\documentclass[12pt]{article}
\usepackage{latexsym}
\usepackage{amsmath}
\usepackage{amssymb}
\catcode`\@=11
\newcount\hour
\newcount\minute
\newtoks\amorpm \hour=\time\divide\hour by 60\minute
=\time{\multiply\hour by 60 \global\advance\minute by-\hour}
\edef\standardtime{{\ifnum\hour<12 \global\amorpm={am}%
        \else\global\amorpm={pm}\advance\hour by-12 \fi
        \ifnum\hour=0 \hour=12 \fi
        \number\hour:\ifnum\minute<10
        0\fi\number\minute\the\amorpm}}
\edef\militarytime{\number\hour:\ifnum\minute<10 0\fi\number\minute}
\def\draftlabel#1{{\@bsphack\if@filesw {\let\thepage\relax
   \xdef\@gtempa{\write\@auxout{\string
      \newlabel{#1}{{\@currentlabel}{\thepage}}}}}\@gtempa
   \if@nobreak \ifvmode\nobreak\fi\fi\fi\@esphack}
        \gdef\@eqnlabel{#1}}
\def\@eqnlabel{}
\def\@vacuum{}
\def\marginnote#1{}
\def\draftmarginnote#1{\marginpar{\raggedright\scriptsize\tt#1}}
\def\draft{
        \pagestyle{plain}
        \overfullrule=2pt
        \oddsidemargin -.1truein
        \def\@oddhead{\sl \phantom{\today\quad\militarytime} \hfil
        \smash{\Large\sl DRAFT} \hfil \today\quad\militarytime}
        \let\@evenhead\@oddhead
        \let\label=\draftlabel
        \let\marginnote=\draftmarginnote
        \def\ps@empty{\let\@mkboth\@gobbletwo
        \def\@oddfoot{\hfil \smash{\Large\sl DRAFT} \hfil}
        \let\@evenfoot\@oddhead}
        \def\@eqnnum{(\theequation)\rlap{\kern\marginparsep\tt\@eqnlabel}%
        \global\let\@eqnlabel\@vacuum}  }

\renewcommand{\theequation}{\thesection.\arabic{equation}}
\renewcommand{\thefootnote}{\fnsymbol{footnote}}
\newcommand{\newsection}{    % Numeration of eqs. is automatic
\setcounter{equation}{0}\section}
\def\appendix#1{\addtocounter{section}{1}\setcounter{equation}{0}
\renewcommand{\thesection}{\Alph{section}}
\section*{Appendix \thesection\protect\indent \parbox[t]{11.15cm}{#1}}
\addcontentsline{toc}{section}{Appendix \thesection\ \ \ #1}}

\def \bi{\bibitem}
\def \la {\label}

\def \b {\beta}

\def\be{\begin{equation}}
\def\ee{\end{equation}}

\newcommand{\cont}[1]{{}_{#1}{}^{#1}}

\catcode`\@=11

\def\bea{\begin{eqnarray}}
\def\eea{\end{eqnarray}}
\def\beann{\begin{eqnarray*}}
\def\eeann{\end{eqnarray*}}
\def\beq{\begin{equation}}
\def\eeq{\end{equation}}
\def\ba{\begin{array}}
\def\ea{\end{array}}
\def\ben{\begin{enumerate}}
\def\een{\end{enumerate}}
 \def \l {\lambda}
 \def\m {\mu}

 \def \la {\label}
 \def\be{\begin{equation}}
\def\ee{\end{equation}}

\def \la {\label}

\font\mybb=msbm10 at 11pt

\def\bb#1{\hbox{\mybb#1}}

\def\bR {\bb{R}}

\def\bH {\bb{H}}
\def\bC {\bb{C}}

\def\e  {\epsilon}

\def \ee {\epsilon}

\def \bi{\bibitem}
\def\a{\alpha }

\def \G {\Gamma}
\def \l {\lambda}
\def \m {\mu}

\def \n {\nu}

\def \LF {{LF}}

\def\be{\begin{equation}}
\def\ee{\end{equation}}

\def \bi {\bibitem}
\def \la{\label}

\begin{document}
%\draft
\date{November 2002}
%%%%%%%%%%%%%%%%%%%%%%%%%%%%%%%%%%%%%%%%%%%%%%%%%%%%%%%%%%
\begin{titlepage}
\begin{center}
%\today
\hfill hep-th/0604079\\
\hfill KUL-TF-06/14\\

\vspace{2.0cm}
{\Large \bf Maximally supersymmetric $G$-backgrounds \\ of IIB supergravity}\\[.2cm]

\vspace{1.5cm}
 {\large  U. Gran$^1$, J. Gutowski$^2$,  G. Papadopoulos$^3$ and D. Roest$^3$}

 \vspace{0.5cm}
${}^1$ Institute for Theoretical Physics \\
K.U. Leuven \\
Celestijnenlaan 200D\\
B-3001 Leuven, Belgium \\
\vspace{0.5cm}
${}^2$ DAMTP, Centre for Mathematical Sciences\\
University of Cambridge\\
Wilberforce Road, Cambridge, CB3 0WA, UK \\
\vspace{0.5cm}
${}^3$ Department of Mathematics\\
King's College London\\
Strand\\
London WC2R 2LS, UK
\end{center}

\vskip 1.5 cm
\begin{abstract}

We classify the geometry of  all supersymmetric  IIB backgrounds which admit
the maximal number of $G$-invariant Killing spinors.
  For compact
stability subgroups $G=G_2, SU(3)$ and $SU(2)$, the spacetime is
locally isometric to a product $X_n\times Y_{10-n}$ with $n=3,4,6$,
where $X_n$ is a maximally supersymmetric solution of a
$n$-dimensional supergravity theory and $Y_{10-n}$ is a Riemannian
manifold with holonomy $G$. For  non-compact stability subgroups,
$G=K\ltimes\bR^8$, $K=Spin(7)$, $SU(4)$, $Sp(2)$, $SU(2)\times
SU(2)$ and $\{1\}$, the spacetime is a pp-wave propagating in an
eight-dimensional manifold with  holonomy $K$. We find new
supersymmetric pp-wave solutions of IIB supergravity.

\end{abstract}
\end{titlepage}
%%%%%%%%%%%%%%%%%%%%%%%%%%%%%%%%%%%%%%%%%%%%%%%
\newpage
\setcounter{page}{1}
\renewcommand{\thefootnote}{\arabic{footnote}}
\setcounter{footnote}{0}

\setcounter{section}{0}
\setcounter{subsection}{0}
%%%%%%%%%%%%%%%%%%%%%%%%%%%%%%%%%%%
\newsection{Introduction}

Supersymmetric backgrounds in supergravity theories can be  characterized by the number of Killing spinors $N$ and
their stability subgroup $G$ in an appropriate spin group \cite{josew}. For a given stability
subgroup $G$, it has been shown in \cite{ggpg2, ggpr} that the Killing spinor equations of IIB supergravity \cite{west, schwarz, howe}
simplify for two classes of backgrounds:  (i) the backgrounds that admit the maximal number of $G$-invariant Killing spinors,
 and (ii) the backgrounds that admit half the maximal number of $G$-invariant Killing spinors.
 In particular the Killing spinors for the former case, the maximally supersymmetric $G$-backgrounds,   can be written as
 \bea
 \epsilon_i=\sum_{j=1}^N\,f_{ij}\,\eta_j~,~~~j=1, \dots, N=2m~,
 \eea
 where $\eta_p$, $p=1, \dots, m$ is a basis of $G$-invariant Majorana-Weyl spinors, $\eta_{m+p}=i\eta_p$,
 and $(f_{ij})$ is a $N\times N$
 matrix with real spacetime functions as entries. In addition, the Killing spinor equations and their
 integrability conditions factorize, see also  appendix A.

The IIB Killing spinors are invariant under the  stability subgroups
$Spin(7)\ltimes\bR^8~ (N=2)$, $SU(4)\ltimes\bR^8~(N=4)$,
$Sp(2)\ltimes\bR^8$ (N=6), $(SU(2)\times SU(2))\ltimes\bR^8~(N=8)$,
$\bR^8~ (N=16)$, $G_2~(N=4)$, $SU(3)~(N=8)$, $SU(2)~(N=16)$ and
$\{1\}~(N=32)$, where $N$ denotes the (maximal) number of invariant
spinors in each case. The maximally supersymmetric IIB backgrounds,
$\{1\}~(N=32)$, have been classified in \cite{ffp}, where it was
found that they are locally isometric to Minkowski spacetime
$\bR^{9,1}$, $AdS_5\times S^5$ \cite{schwarz} and the Hpp-wave
\cite{bfhpwave} . In addition, the geometry of the maximally
supersymmetric $Spin(7)\ltimes\bR^8$-, $SU(4)\ltimes\bR^8$- and
 $G_2$-backgrounds
  has already been investigated  \cite{ggpg2, ggpr} using the spinorial geometry method of \cite{ggpm}. Here we shall use the same method
  to investigate the remaining cases. There are two classes of maximally supersymmetric $G$-backgrounds depending on whether
  $G$ is a compact or non-compact subgroup of $Spin(9,1)$. The geometry of the backgrounds in the two cases is distinct.
To outline our results we denote with    $ds^2(S^k)$  the metric of
the round $k$-dimensional sphere $S^k$, with $ds^2(AdS_k)$ the
metric of $k$-dimensional anti-de-Sitter space $AdS_k$ and with
$ds^2(CW_k(A))$ the metric\footnote{ The metric is $ds^2(CW_k(A))=2
dx^- (dx^++ \tfrac{1}{2} A_{ij} x^i x^j\, dx^-)+ (dx^i)^2$, see
\cite{cahen}.}
 of the $k$-dimensional Cahen-Wallach space $CW_k(A)$ associated with the (constant) quadratic form $A$.
The metric and fluxes are expressed in terms of  orthonormal or  null frame bases which arise from the description
of the spinors in terms of forms. Our spinor conventions can be found  in \cite{ggpr}.

\subsection{Backgrounds with compact stability subgroups}

The geometry of the maximally supersymmetric $G$-backgrounds, where
$G$ is a compact subgroup of $Spin(9,1)$, is as follows:

\begin{itemize}

\item $G_2$: The spacetime is locally isometric to the product $\bR^{2,1}\times Y_7$, where $Y_7$ is a $G_2$ holonomy manifold.
The metric and fluxes  are
\bea
ds^2(M)=ds^2(\bR^{2,1})+ds^2(Y_7)~,~~~G=P=F=0~,
\eea
i.e. the fluxes vanish.

\item $SU(3)$: The spacetime $M$ is locally isometric to a product of a four-dimensional symmetric Lorentzian space
and a six-dimensional Calabi-Yau manifold $Y_6$. In particular, the spacetime is
\begin{itemize}
\item $M=AdS_2\times S^2\times Y_6$, and the metric and fluxes are
\bea &&ds^2(M)=ds^2(AdS_2)+ds^2(S^2)+ ds^2(Y_6)~, \cr
&&ds^2(AdS_2)=-(e^0)^2+(e^1)^2~,~~~ds^2(S^2)=(e^5)^2+(e^6)^2~, \cr
&&F= {1 \over 2\sqrt{2}}[H^1 \wedge {\rm Re}\chi- H^2\wedge {\rm Im}
\chi]~,~~~\chi=(e^2+ie^7)\wedge (e^3+ie^8)\wedge (e^4+ie^9)~, \cr
&&H^1=\lambda_1\, e^0\wedge e^1+\lambda_2\, e^5\wedge
e^6~,~~~H^2=-\lambda_1\, e^5\wedge e^6+\lambda_2\, e^0\wedge e^1~,
\cr &&G=P=0~, \eea where the scalar  curvature of $AdS_2$ and $S^2$
are $R_{AdS_2}=-R_{S^2}=-4(\l_1^2+\l_2^2)$.

\item $M=CW_4(-2\m^2 {\bf 1})\times Y_6$, and the metric and fluxes are
\bea &&ds^2(M)=ds^2(CW_4)+ ds^2(Y_6)~, \cr &&F={1 \over
2\sqrt{2}}[H^1 \wedge {\rm Re}\chi- H^2\wedge {\rm Im} \chi]~, \cr
&&H^1=\m\, e^-\wedge e^1~,~~~H^2=\m\, e^-\wedge e^6~, \cr &&G=P=0\,.
\eea
\item $M=\bR^{3,1}\times Y_6$, and the metric and fluxes are
\bea
&&ds^2(M)=ds^2(\bR^{3,1})+ ds^2(Y_6)~,
\cr
&&F=G=P=0~.
\eea

\end{itemize}
\item $SU(2)$: The spacetime $M$ is locally isometric to a product of a six-dimensional symmetric Lorentzian space
and a four-dimensional hyper-K\"ahler manifold $Y_4$. In particular, the spacetime is
\begin{itemize}
\item $M=AdS_3\times S^3\times Y_4$, and the metric and fluxes are
\bea &&ds^2(M)=ds^2(AdS_3)+ds^2(S^3)+ ds^2(Y_4)~, \cr
&&ds^2(AdS_3)=-(e^0)^2+(e^1)^2+(e^2)^2~,~~~ds^2(S^3)=(e^3)^2+(e^4)^2+(e^5)^2~,
\cr &&F= \tfrac{1}{4} v\cdot \hat \omega \wedge H~, \cr &&G=(v^4+i
v^5) H~,~~~H=\l\, e^0\wedge e^1\wedge e^2+\l\, e^3\wedge e^4\wedge
e^5~, \cr &&P=0~, \eea where $v\cdot\hat
\omega=v^1\hat\omega_I+v^2\hat\omega_J+v^3\hat\omega_K$ is a linear
superposition of the K\"ahler forms $\hat\omega_I, \,\hat\omega_J$
and $\hat\omega_K$ of the hyper-K\"ahler manifold $Y_4$, $v^2=1$ and
the scalar curvature $R_{AdS_3}=-R_{S^3}=-\tfrac{3}{2}\l^2$.

\item $M=CW_6(-\tfrac{1}{4}\mu^2 {\bf 1})\times Y_4$, and the metric and fluxes are
\bea &&ds^2(M)=ds^2(CW_6)+ ds^2(Y_4)~, \cr &&F=\tfrac{1}{4} v\cdot
\hat \omega \wedge H~, \cr &&G=(v^4+i v^5) H~,~~~H=\m e^-\wedge
e^1\wedge e^2-\m e^-\wedge e^6\wedge e^7~, \cr &&P=0~. \eea
\item $M=\bR^{5,1}\times Y_4$, and the metric and fluxes are
\bea
&&ds^2(M)=ds^2(\bR^{5,1})+ ds^2(Y_4)~,
\cr
&&F=G=P=0~.
\eea
\end{itemize}

\end{itemize}

Therefore, we have shown that the maximally supersymmetric $G_2$-, $SU(3)$-, and $SU(2)$-backgrounds for $G$ compact
 are the maximally supersymmetric solutions of ${\cal N}=1$, ${\cal N}=2$ and (2,0)-supergravities in three, four and six dimensions, respectively, lifted
 to IIB supergravity.
The maximally supersymmetric solutions for the ${\cal N}=2$
four-dimensional supergravity have been found in \cite{kgd4}, see
\cite{ortin} for a more recent account. In six dimensions, the
maximally supersymmetric solutions of (1,0) supergravity have been
classified in \cite{gmr} and of the (2,0) supergravity in
\cite{jose}. In three dimensions, it is straightforward to show that
the only maximally supersymmetric solution is locally isometric to
Minkowski spacetime.

\subsection{Backgrounds with non-compact stability subgroups}
Next we turn to investigate the geometry of maximally supersymmetric $G=K\ltimes\bR^8$-backgrounds for
$K=Spin(7), SU(4), Sp(2), SU(2)\times SU(2)$ and $\{1\}$.
It turns out that the spacetime $M$ always admits a null parallel vector field $X$ and the holonomy of the Levi-Civita
connection of spacetime is contained in $K\ltimes\bR^8$, i.e.
\bea
\nabla_A X=0~,~~~{\rm hol}(\nabla)\subseteq K\ltimes\bR^8~.
\eea
Therefore, the spacetime
is a pp-wave propagating in an eight-dimensional Riemannian
 manifold
$Y_8$ such that ${\rm hol}(\tilde \nabla)\subseteq K$, where $\tilde \nabla$ is the Levi-Civita connection of $Y_8$.
Alternatively, the spacetime is a two-parameter Lorentzian  deformation family of $Y_8$. Adapting
coordinates along the parallel vector field $X=\partial/\partial u$, the metric can be written as
\bea
ds^2=2 dv (du+Vdv+n)+ds^2(Y_8)=2 dv (du+Vdv+n)+\gamma_{IJ} dy^I dy^J~,~
\la{metrnull}
\eea
where the metric $\gamma_{IJ}=\delta_{ij} e^i_I e^j_J$ of $Y_8$ may also depend on the coordinate $v$.
The requirement that ${\rm hol}(\tilde \nabla)\subseteq K$
implies that the components $ e^A \,\Omega_{A, ij} $ of the connection one-form take values in the Lie algebra
of $K$, $\mathfrak{k}$.

In all cases, the fluxes are null, i.e. \bea P=P_-(v)
e^-~,~~~G=e^-\wedge L~,~~~F=e^-\wedge M~, \la{fluxLM} \eea and the
Bianchi identities give $dP=dG=dF=0$, where $L$ and $M$ are a two-
and a self-dual  four-form, respectively, of $Y_8$. In particular,
one finds that $P_-=P_-(v)$. The most convenient way to give the
conditions that the Killing spinor equations impose on the fluxes is
to decompose $L\in \Lambda^2(\bR^8)\otimes \bC$  and $M\in
\Lambda^{4+} (\bR^8)$ in irreducible representations of $K$. In
particular, one finds that \bea L=L^{\mathfrak{k}}+L^{\rm
inv}~,~~~M= M^{\rm inv}+\tilde M~, \la{fluxLMinv} \eea where
$L^{\mathfrak{k}}$ is the Lie algebra valued component of $L$
 in the decomposition $\Lambda^2(\bR^8)=\mathfrak{k}+\mathfrak{k}^\perp$,
and $L^{\rm inv}$ and $M^{\rm inv}$ are $K$-invariant two- and
four-forms, respectively. $M^{\rm inv}$ decomposes further as
$M^{\rm inv}=m^0+\hat M^{\rm inv}$, where $m^0$ has the property
that the associated Clifford algebra element satisfies
$m^0\epsilon=g \epsilon$, $g\not=0$ a spacetime function, for all
Killing spinors $\epsilon$. In a particular gauge, the Killing
spinor equations imply that $g$ is proportional to $Q_-$
 and  restrict   the spacetime dependence
of  $L^{\rm inv}$ and
 $M^{\rm inv}$.
Furthermore,  $\tilde M$ takes values in a representation of $K$ in
$\Lambda^{4+} (\bR^8)$ with the property that the associated
Clifford algebra element satisfies $\tilde M\epsilon=0$ for all
Killing spinors $\epsilon$. $L^{\mathfrak{k}}$ and $\tilde M$ are
not determined by the Killing spinor equations.
   In particular, one finds\footnote{To solve all conditions that arise from the Killing spinor
   equations, we present our results in a particular gauge. }    the following:

\begin{itemize}
\item $Spin(7)\ltimes\bR^8$:
\bea G=e^-\wedge L^{\mathfrak{spin}(7)}~,~~~F=e^-\wedge
(\tfrac{1}{14} Q_-(v) \psi+ M^{\bf 27})~, \eea where $\psi$ is the
invariant $Spin(7)$ four-form, $Q_-$ depends only on $v$,
 and $L^{\mathfrak{spin}(7)}$ and $\tilde M=M^{\bf 27}$
are not determined in terms of the geometry.

\item $SU(4)\ltimes\bR^8$:
\bea &&G=e^-\wedge (L^{\mathfrak{su}(4)}+ \ell(v) \omega)~, \cr
&&F=e^-\wedge \big(- \tfrac{1}{12} Q_-(v) \omega\wedge \omega+{\rm
Re}\,(m(v)\, \chi)+ \tilde M^{2,2}\big) \la{cysu4} \eea where $\chi$
is the $SU(4)$-invariant (4,0)-form, $\ell$, $m$ and $Q_-$ depend
only on $v$ as indicated, and $\tilde M=\tilde M^{2,2}$ is a
traceless (2,2)-form.

\item $Sp(2)\ltimes\bR^8$:
\bea && G=e^-\wedge (L^{\mathfrak{sp}(2)}+ \ell^r(v)\,
\omega_r)~,~~~ \cr && F=e^-\wedge (-\tfrac{1}{20}Q_-(v) \psi+
m^{rs}(v) \omega_r\wedge \omega_s+ M^{{\bf 14}}) \eea where
$\omega_I=\omega_1$, $\omega_J=\omega_2$ and $\omega_K=\omega_3$ are
the Hermitian forms of the quaternionic endomorphisms $I$, $J$ and
$K$,
 $\psi=\sum_{r=1}^3 \omega_r\wedge \omega_r$, $m^{rs}$ is a symmetric traceless $3\times 3$-matrix
 that depends only on $v$, $\ell^r=\ell(v)$,
and $\tilde M=M^{{\bf 14}}$.

\item $(SU(2)\times SU(2))\ltimes\bR^8$:
\bea &&G=e^-\wedge (L^{\mathfrak{su}(2)\oplus \mathfrak{su}(2)}
+\ell^1(v)\omega_1+ \ell^2(v) \omega_2+\ell^3(v) \chi_1 \cr
&&~~~~+\ell^4(v) \chi_2+\ell^5(v) \bar\chi _1+\ell^6(v)\bar\chi_2
)~,~~~ \cr &&F=e^-\wedge (-\tfrac{1}{4} Q_-(v)[\omega_1\wedge
\omega_1+ \omega_2\wedge \omega_2] +m^1(v)\omega_1\wedge \omega_2+
{\rm Re}[m^2(v) \omega_1\wedge \chi_2 \cr &&~~~~+m^3(v)
\omega_2\wedge \chi_1 +m^4(v) \chi_1\wedge \chi_2+m^5(v)
\chi_1\wedge \bar\chi_2]+ M^{({\bf 3}, {\bf3})})~, \eea where the
pairs $(\omega_1, \chi_1)$ and $(\omega_2, \chi_2)$ are the
hermitian (1,1)- and holomorphic (2,0)-forms
 associated with the $(SU(2)\times SU(2))\ltimes\bR^8$-structure,  $\ell, m$ depend only on $v$,
  and $\tilde M=M^{({\bf 3}, {\bf3})}$.

\item $\bR^8$:
\bea
G=e^-\wedge L(v)~,~~~F=e^-\wedge M(v)~,
\eea
where $L$ and $M$ are a two- and a self-dual four-form on $\bR^8$, respectively, and depend
only on $v$.

\end{itemize}

The integrability conditions of the Killing spinor equations and the Bianchi identities imply
that all field equations are satisfied provided that $E_{--}=0$, where $E_{--}$ denotes the
$--$ component of the Einstein equations. This in turn gives
\bea
  && - (\partial^i + \Omega_{j,}{}^{ji})(\partial_i V -
  \partial_v n_I e^I{}_i) + \tfrac{1}{4} (dn)_{ij} (dn)^{ij} -
  \tfrac{1}{2} \gamma^{IJ} \partial_v{}^2 \gamma_{IJ}
  - \tfrac{1}{4} \partial_v \gamma^{IJ} \partial_v \gamma_{IJ} \nonumber \\
  &&
 - \tfrac{1}{6} F_{- i_1 \cdots i_4} F_-{}^{i_1 \cdots i_4}
  - \tfrac{1}{4} G_{-}{}^{i_1 i_2} G^*_{- i_1 i_2} -2 P_- P_-^* = 0 \,,
 \label{Einstein-eq}
 \eea
where $\gamma^{IJ}$ is the inverse of the metric $\gamma_{IJ}$
defined in ({\ref{metrnull}}). For the special
case of  fields  independent of $v$, this equation
becomes
 \bea
  - \Box_8 V + \tfrac{1}{4} (dn)_{ij} (dn)^{ij} - \tfrac{1}{6} F_{- i_1 \cdots i_4} F_-{}^{i_1 \cdots i_4}
  - \tfrac{1}{4} G_{-}{}^{i_1 i_2} G^*_{- i_1 i_2} -2 P_- P_-^* = 0 \,,
 \label{Einstein-eq-trunc}
 \eea
where $\Box_8$ is the Laplacian on the eight-dimensional space $Y_8$ and
$dn$  takes values in $\mathfrak{k}$.

The backgrounds that we have found can be thought of as vacua of IIB
string theory. This particulary applies to compact stability
subgroups. The backgrounds  $\bR^{9-n, 1}\times Y_n$ are vacua of
IIB compactifications on $G_2$ for $n=7$, and on Calabi-Yau
manifolds for $n=6$ and $n=4$. The backgrounds $AdS_{5-n/2}\times
S^{5-n/2}\times Y_n$ can be thought either of as the vacua of the
Calabi-Yau or $S^{5- n/2}\times Y_n$ compactifications with fluxes.
For a recent application of the latter see \cite{verlinde}. For
non-compact stability subgroups, the situation is different. If one
views the solutions as vacua of compactifications and so insists to
be invariant under lower-dimensional Poincar\'e symmetry, then the
only solutions are $\bR^{9-n,1}\times Y_n$. In particular all the
fluxes vanish because of the field equations.

This paper is organized as follows: In sections two and three, we describe the geometry of maximally supersymmetric $SU(3)$- and $SU(2)$-backgrounds,
respectively.
In section four, five and six, we give present the maximally supersymmetric $Sp(2)\ltimes\bR^8$-, $(SU(2)\times SU(2))\ltimes\bR^8$-
and $\bR^8$-backgrounds, respectively. In section seven, we describe solutions of maximally supersymmetric
$G$-backgrounds, for a non-compact $G$. In appendix A, we summarize the Killing spinor equations
and some of their integrability conditions.

\newsection{Maximal $SU(3)$-backgrounds}

\subsection{ Supersymmetry conditions}

As we have mentioned in the introduction to solve the Killing spinor equations  and the
integrability conditions
of maximally
supersymmetric $SU(3)$-backgrounds,
one may use a basis in the Majorana-Weyl
 $SU(3)$-invariant spinors of IIB supergravity. Such a  basis\footnote{
 Note that the $SU(3)$-invariant spinors are annihilated by
$\Gamma^{\mu_1 \bar \mu_2}$, where $\mu_1 \neq \mu_2$: indeed this
gives rise to two independent projection operators, allowing for
eight supersymmetries.}
  is
\bea
\eta_1&=& 1+e_{1234}~,~~~\eta_2= i (1-e_{1234})~, \cr \eta_3&=&
e_{15}+e_{2345}~,~~~\eta_4= i (e_{15}-e_{2345})~.
\eea
To proceed, it is convenient to introduce the notation
 $A
= (a, m)$. Here $a = (\a, \bar \a)$, $\a = (-,1)$ and $\bar \a = (+,
\bar 1)$ are the `world-volume' labels and $m = (\mu, \bar \mu)$,
$\mu = (2,3,4)$ and $\bar \mu = (\bar 2, \bar 3, \bar 4)$ denote
those of the `transverse space'. Due to the null directions,   $X^{\bar \a}\not= (X^\a)^*$
for a real vector field $X$.

The algebraic Killing spinor equations (\ref{facalg}) imply that all
components of the $P$-flux vanish. In addition, the same equation requires that
 \bea
   && G_{\mu_1 \mu_2 \mu_3} = G_{\mu_1} \cont{\mu_2} = G_{\bar \mu_1} \cont{\mu_2} = G_{\bar \mu_1 \bar \mu_2 \bar \mu_3} = 0 \,, \notag \\
   && G_{\a \mu_1 \mu_2} = G_{\a_1} \cont{\mu} - G_{\a_1} \cont{\a_2} = G_{\a \bar \mu_1 \bar \mu_2} = 0 \,,\notag \\
   && G_{\bar \a \mu_1 \mu_2} = G_{\bar \a_1} \cont{\mu} + G_{\bar \a_1} \cont{\a_2} = G_{\bar \a \bar \mu_1 \bar \mu_2} = 0 \,,\notag \\
   && G_{\a_1 \bar \a_2 \mu} - \tfrac{1}{2} g_{\a_1 \bar \a_2} G_{\mu} \cont{\a_3} = G_{\a_1 \a_2 \bar \mu} = G_{\bar \mu} \cont{\a} = G_{\bar \a_1 \bar \a_2 \bar \mu} = 0 \,.
 \eea
The gravitino Killing spinor equations (\ref{facpar}) involving $G$ imply
 \bea
  G_{A b m} = G_{A \mu_1 \mu_2} = G_{A \bar \mu_1 \bar \mu_1} = 0 \,.
 \eea
Combining the above results from the gravitino and algebraic Killing spinor equations,
one finds that
\bea
P=G=0~,
\la{gpconsu3}
\eea
i.e. all the $P$ and $G$ fluxes vanish.

The gravitino Killing spinor equations require that $F$ satisfies
 \bea
  && F_{A \mu_1 \mu_2} \cont{\mu_3} = 0 \,, \notag \\
  && F_{A \a_1 \mu_1} \cont{\a_2} - F_{A \a_1 \mu_1} \cont{\mu_2} = F_{A \bar \a_1 \mu_1} \cont{\a_2} + F_{A \bar \a_1 \mu_1} \cont{\mu_2} = 0 \,, \notag \\
  && F_{A \mu_1 \mu_2 \a_1 \bar \a_2} - \tfrac{1}{2} g_{\a_1 \bar \a_2} F_{A \mu_1 \mu_2} \cont{\a_3} = 0 \,,
 \eea
from which follows that
 \bea
  F_{\mu_1 \mu_2 \mu_3 \bar \mu_4 \bar \mu_5} = F_{a \mu_1 \mu_2 \mu_3 \bar \mu_4} = F_{a_1 a_2 a_3 \mu_1 \mu_2} = 0~.
 \eea
Subsequently, the self-duality constraint on $F$ implies that
 \bea
  F_{a \mu_1 \mu_2 \bar \mu_3 \bar \mu_4} = F_{a_1 a_2 \mu_1 \mu_2 \bar \mu_3} = F_{a_1 a_2 a_3 \mu_1 \bar \mu_2} = F_{a_1 \cdots a_4 \mu} = 0 \,.
 \eea
Therefore the non-vanishing components of $F$ are
 \bea
  F_{\a_1 \a_2 234} \,, \; F \cont{\a} {}_{234} \,, \; F_{\bar \a_1
  \bar \a_2 234} \,, \; \tilde{F}_{\a_1 \bar \a_2 \bar 2 \bar 3 \bar
  4} \,,
 \label{max-SU(3)-flux}
 \eea
and their complex conjugates, where tilde denotes the traceless part. These are all singlets under self-duality.

Next turn to the conditions on the geometry, the equation (\ref{facpar}) implies the constraints
 \bea
  \Omega_{A, b m} = \Omega_{A, \mu_1 \mu_2} = 0 \,,
  \la{geoconsu3}
 \eea
for the spin connection.

The remaining components of the spin connection and fluxes give rise to the following parallel
transport equation:
 \bea
  &&  \partial_A \epsilon - \tfrac{1}{2} i Q_A\epsilon + \tfrac{1}{2} \Omega_{A,} \cont{\mu} \G^{2 \bar 2}\epsilon
  + \tfrac{1}{4} \Omega_{A,b_1 b_2} \G^{b_1 b_2}\e + \tfrac{1}{2} i F_{A b 234}
  \Gamma^{b 234}\e +
  \notag \\
   && \; \; \; + \tfrac{1}{2} i F_{A b \bar 2 \bar 3 \bar 4}
  \Gamma^{b \bar 2 \bar 3 \bar 4}\e= 0 \,.
 \label{max-SU(3)-connection}
 \eea
The generators $1$, $\Gamma^{2 \bar 2}$ and $\G^{b_1 b_2}$ span a
$\mathfrak{u}(1)^2 \oplus \mathfrak{so}(3,1)$ algebra inside $\mathfrak{u}(1) \oplus \mathfrak{spin}(9,1)$.
However, for $A=a$ there are also the generators $i \Gamma^{b 234}$
and $i \Gamma^{b \bar 2 \bar 3 \bar 4}$ in this connection due to
the non-vanishing flux components (\ref{max-SU(3)-flux}). Note that
these generators satisfy the same algebra as $1$, $\Gamma^{2 \bar
2}$, $\G^{b_1 b_2}$, $\G^{b 2}$ and $\G^{b \bar 2}$; therefore the
connection in (\ref{max-SU(3)-connection}) takes values in a $\mathfrak{u}(1)
\oplus \mathfrak{so}(5,1) \equiv \mathfrak{u}(1) \oplus \mathfrak{sl}(2,\mathbb{H})$ algebra\footnote{
Note that the holonomy of the supercovariant connection of ${\cal N}=2$ ungauged supergravity
in four dimensions is $SL(2,\bH)$ \cite{bat}.} which is
not embedded in the $\mathfrak{u}(1) \oplus \mathfrak{spin}(9,1)$ gauge symmetry. Because of this,
one cannot set the connection to zero by a suitable gauge
transformation. Observe that the traceless part of
$\Omega_{A, \mu_1 \bar \mu_2}$ does not appear in the parallel transport equations.

The vanishing of the curvature of the connection appearing in
(\ref{max-SU(3)-connection}) gives rise to the following equations:
 \bea
  && \partial_{[ A} Q_{B ]} = 0 \,,~~~ R_{AB,} \cont{\mu}
  - 8 F_{[A | c 234} F_{B ]}{}^c {}_{\bar 2 \bar 3 \bar 4} = 0 \,, \cr
  && R_{A B, c_1 c_2} - 8 F_{[A| c_1 234} F_{B] c_2 \bar 2 \bar 3 \bar 4}
  +8 F_{[A| c_2 234} F_{B] c_1 \bar 2 \bar 3 \bar 4} = 0 \,, ~~~
  \nabla_{[A} F_{B] c 234} = 0 \,.
  \la{intconsu3}
 \eea
It will be important in the following that the flux bilinear terms in the first line vanish due to the
conditions (\ref{max-SU(3)-flux}) on $F$.
The conditions (\ref{gpconsu3}), (\ref{geoconsu3}), (\ref{max-SU(3)-flux}) and
(\ref{intconsu3}) impose restrictions on the geometry of spacetime which we shall investigate.

\subsection{Geometry of spacetime}

We write the spacetime metric as $ds^2=\eta_{ab} e^a e^b+\delta_{mn}
e^m e^n$. The torsion free condition for the frame $e^a, e^m$ and
the condition $\Omega_{A,bm}=0$ in (\ref{geoconsu3}) imply that the
spacetime admits an integrable bi-distribution of co-dimensions four
or six, i.e. both $\{ e^a\}$ and $\{ e^m\}$ span an integrable
distribution. Therefore the spacetime $M$ is locally a topological
product, $M=X_4\times Y_6$. Furthermore,  $\Omega_{A,bm}=0$ in
(\ref{geoconsu3}) implies that the metric compatible product
structure $\pi=\eta_{ab} e^a e^b-\delta_{mn} e^m e^n$ is parallel
with respect to the Levi-Civita connection. This in turn implies
that $\pi$ is integrable and in the coordinate system that $\pi$ is
diagonal, the metric is a product. In particular, \bea
ds^2(M)=ds^2(X_4)+ds^2(Y_6)~,~~~~ds^2(X_4)=\eta_{ab} e^a
e^b~,~~~ds^2(Y_6)=\delta_{mn} e^m e^n~, \eea i.e. $ds^2(X_4)$ does
not depend on the coordinates of $Y_6$ and vice-versa. The geometry
of $X_4$ and $Y_6$ can be separately investigated. First consider
the geometry of $Y_6$. The condition $\Omega_{m,\mu_1\mu_2}=0$ in
(\ref{geoconsu3}) and $\Omega_{m,\mu}{}^\mu=0$,
 which can be easily derived from (\ref{intconsu3}) after a suitable choice of gauge, imply
that $Y_6$ is Calabi-Yau. There are no additional conditions on $Y_6$.

Next let us turn to investigate the geometry of $X_4$. For this,
observe that the five form can be written as \bea F= {1 \over
2\sqrt{2}}[H^1 \wedge {\rm Re}\chi- H^2\wedge {\rm Im} \chi]~, \eea
where $\chi$ is the parallel (3,0)-form on the Calabi-Yau manifold
$Y_6$, and $H^1$ and $H^2$ are two-forms on $Y_4$. In addition the
Bianchi identity of $F$ together with the last equation of
(\ref{intconsu3}) imply that $H^1, H^2$ are independent of the
coordinates of $X_6$ and are parallel forms on $X_4$. The remaining
conditions can now be written as restrictions on the geometry of
$X_4$. In particular, one has
  \bea
  && R_{a_1a_2, b_1 b_2} - 4H^1_{[a_1| b_1 |} H^1_{a_2] b_2 }-
   4H^2_{[a_1| b_1|} H^2_{a_2] b_2 } = 0 \,, ~~~
\cr &&\nabla_a H^1_{bc} = 0~,~~~\nabla_a H^2_{bc} =
0~,~~~H^1_{[a|c|} H^2_{b]}{}^c=0~,~~~\star H^1=H^2,
\la{yconsu3}
\eea
where the last condition is implied by the self-duality of $F$.
Since $H^1$ and $H^2$ are parallel, the first equation implies that
the Riemann curvature $R$ of $X_4$ is also parallel. Therefore $X_4$
is a Lorentzian symmetric space. The fields $H^1$ and $H^2$ are
uniquely determined by their values at the origin of the symmetric
space up to  rigid $SO(3,1)$ transformations. Since $H^1$ and $H^2$
are related by the Hodge star operator in $X_4$, it suffices to find
$H^1$. It turns out that $H^1$ can be chosen as, see e.g. \cite{ffp, ross},
\bea
\lambda_1\,
e^0\wedge e^1+\lambda_2\, e^5\wedge e^6~,~~~\m\, e^-\wedge e^1~,~~~
\eea and so $H^2$ is
\bea
-\lambda_1\, e^5\wedge e^6+\lambda_2\,
e^0\wedge e^1~,~~~\m\, e^-\wedge e^6~,
\eea
where $\m, \lambda_1$
and $\lambda_2$ are real constants. Therefore $H^1$ defines a
two-plane at the origin of the symmetric space $Y_4$ which is either
time-like and/or space-like, or null. Moreover $H^1$ commutes with
$H^2$. It is straightforward to see that in the case of the
time-like and/or spacelike plane, $X_4=AdS_2\times S^2$, where both
factors have the same radius and scalar curvature
$R_{AdS_2}=- 4(\l_1^2+\l_2^2)$ and $R_{S^2}=
4(\l_1^2+\l_2^2)$, respectively.
 In the case that the plane is null $X_4=CW_4(-2\m^2 {\bf 1})$. Note that the three
 different geometries of $X_4$ are related by Penrose limits of $AdS_2\times S^2$ \cite{bhfp}.
These are the maximally supersymmetric solutions of four-dimensional ${\cal N}=2$ supergravity \cite{kgd4}.
This completes the proof for the maximally
 supersymmetric $SU(3)$-backgrounds. The result is summarized in the introduction.

\newsection{Maximal $SU(2)$-backgrounds}

\subsection{Supersymmetry conditions}

 A basis in the space of the $SU(2)$-invariant Majorana-Weyl spinors is
\bea
&&\eta_1= 1+e_{1234}\,,~~\eta_2= i (1-e_{1234})\,,~~\eta_3=
e_{12}-e_{34}\,,
~~\eta_4= i(e_{12}+e_{34})\,,
\cr
&&\eta_5= e_{15}+
e_{2345}\,,~~ \eta_6= i (e_{15}- e_{2345})\,,~~ \eta_7= e_{52}+
e_{1345}\,,~~\eta_8=i (e_{52}- e_{1345})\,.~~ \eea To find the
conditions that the Killing spinor equations of appendix A impose on
the geometry of spacetime, it is convenient to split up the
ten-dimensional frame indices into $A = (a, m)$, where $a = (\a,
\bar \a)$, with $\a = (-,1,2)$ and $\bar \a = (+, \bar 1, \bar 2)$,
and $m = (\mu, \bar \mu)$, with $\mu = (3,4)$ and $\bar \mu = (\bar
3, \bar 4)$.

The algebraic Killing spinor equations (\ref{facalg}) imply that
\bea
P=0~.
\eea
In addition, $G$ is constrained as
 \bea
   && G_{a_1 a_2 \mu} = G_{a_1 a_2 \bar \mu} = G_{a \mu_1 \mu_2}
   = G_{a \mu}{}^{\mu} =
   G_{a \bar \mu_1 \bar \mu_2} = G_{\mu_1 \mu_2 \bar \mu_3} = G_{\mu_1 \bar \mu_2 \bar \mu_3} = 0
\cr
 && \tilde{G}_{\a_1 \a_2 \bar \a_3} = G_{\bar \a_1} \cont{\a_2} =
  G_{\bar \a_1 \bar \a_2 \bar \a_3} = 0 \,,
 \label{M3-constraint}
 \eea
 where tilde denotes the traceless component.
The gravitino Killing spinor equations
(\ref{facpar}) imply that
 \bea
  G_{A b m} = 0 \,.
 \eea
Due to these constraints, the components $G_{a \mu_1 \bar \mu_2}$
also vanish and one is only left with $G_{a_1 a_2 a_3}$ components,
subject to (\ref{M3-constraint}). Incidentally, $G^*$ satisfy the
same conditions as it can be seen by taking the complex conjugate of
those for $G$.

The gravitino Killing spinor equations (\ref{facpar}) together with the self-duality of $F$ imply that
the only non-vanishing components are
 \bea
  F_{b_1 b_2 b_3 \mu_1 \mu_2 } \,, \quad
  F_{b_1 b_2 b_3} \cont{\mu} \,, \quad
  F_{b_1 b_2 b_3 \bar \mu_1 \bar \mu_2 } \,,
  \label{P2-constraint}
 \eea
subject to the conditions \bea \tilde{F}_{\a_1 \a_2 \bar
\a_3\mu_1\mu_2} = F_{\bar \a_1} \cont{\a_2}{}_{\mu_1\mu_2} =
  F_{\bar \a_1 \bar \a_2 \bar \a_3\mu_1\mu_2}=0
  \la{selfdf}
  \eea
  and similarly for the remaining two components. In addition, (\ref{facpar}) requires that the component
 \bea
  && \Omega_{A, b m} = 0 \,,
  \la{geoconsu2}
 \eea
of the spin connection.

Using the above conditions on the fluxes, the parallel transport equation becomes
 \bea
  && \partial_A \e - \tfrac{1}{2} i Q_A\e
  + \tfrac{1}{4} \Omega_{A,b_1 b_2} \G^{b_1 b_2} \e+ \tfrac{1}{2} \Omega_{A, 34} \G^{34}\e +
  \tfrac{1}{2} \Omega_{A,} \cont{\mu}
  \G^{3 \bar 3}\e + \tfrac{1}{2} \Omega_{A, \bar 3 \bar 4} \G^{\bar 3 \bar 4}\e + \notag \\
  && + \tfrac{i}{8} F_{A b_1 b_2 m_1 m_2} \G^{b_1 b_2 m_1 m_2}\e + \tfrac{1}{8} G_{A b_1 b_2} \G^{b_1 b_2}
  C*\e = 0 \,.
 \label{SU(2)-pte}
 \eea
A necessary condition for the existence of solutions to this parallel transport equation is the vanishing
of the curvature. This leads to the conditions
 \bea
  && \partial_{[A} Q_{B]} - \tfrac{1}{16} i G_{[A| c_1 c_2} G^*_{B]}{}^{c_1 c_2} =0 \,, \cr
  && R_{AB,34} - F_{[A| b_1 b_2 3 Q} F_{B]}{}^{b_1 b_2} {}_{4} {}^Q = 0 \,,
   \cr
  && R_{AB,} \cont{\mu} -F_{[A| b_1 b_2 \mu n} F_{B]}{}^{b_1 b_2 \mu n} = 0 \,,
\cr
  && R_{AB, b_1 b_2} - \tfrac{1}{4} G_{[A| b_1 c} G^*_{B] b_2}{}^c + \tfrac{1}{4} G_{[A| b_2 c} G^*_{B] b_1}{}^c
    - 2 F_{[A| b_1 c m_1 m_2} F_{B] b_2}{}^{c m_1 m_2} = 0 \,, \cr
  && \nabla_{[A} F_{B] b_1 b_2 m_1 m_2} = (\nabla_{[A} - i Q_{[A}) G_{B] b_1 b_2} = 0 \,,
  \cr
&& F^{[A}{}_{  m_1 n [b_1 b_2} F^{B]}{}_{ b_3 b_4] m_2}{}^n
-F^{[A}{}_{  m_2 n [b_1 b_2} F^{B]}{}_{ b_3 b_4] m_1}{}^n=0 \cr
&&F^{[A}{}_{ m_1 m_2[ b_1 b_2} G^{B]}{}_{ b_3 b_4]} = G^{[A}{}_{[
b_1 b_2} (G^*)^{B]}{}_{ b_3 b_4]} = 0 \,,
 \label{su2curv}
 \eea
The flux bilinear terms in the first three lines vanish due to the conditions (\ref{M3-constraint}) and (\ref{selfdf}).
It remains to solve these conditions and find the geometry of spacetime.

\subsection{Geometry of spacetime}

The metric of the spacetime can be written as $ds^2=\eta_{ab} e^a
e^b+\delta_{mn} e^m e^n$. In addition (\ref{geoconsu2}) implies that
the spacetime $M$ admits an integrable bi-distribution of
co-dimension six and a metric compatible parallel product structure
$\pi$. As in the $SU(3)$ case previously,
 $M=X_6\times Y_4$, where $X_6$ is a
Lorentzian manifold and $Y_4$ is a Riemannian manifold. In addition,  the metric is  a
product, i.e.
\bea
ds^2(M)=ds^2(X_6)+ds^2(Y_4)~,~~~ds^2(X_6)=\eta_{ab} e^a
e^b~,~~~ds^2(Y_4)=\delta_{mn} e^m e^n~,
\eea
where $ds^2(X_6)$ does not dependent on the coordinates of $Y_4$ and
vice-versa. First let us examine the geometry of $Y_4$. It is
straightforward to observe from (\ref{su2curv}) that the components
$R_{mn,\mu}{}^\mu$ and $R_{mn,34}$ of the Riemann curvature vanish.
These curvature components span an $\mathfrak{su}(2)$ subalgebra in
$\mathfrak{so}(4)=\mathfrak{su}(2)\oplus \mathfrak{su}(2)\subset
\mathfrak{spin}(9,1)$. This implies that the holonomy of the
Levi-Civita connection of $Y_4$ is contained in $SU(2)$ and so $Y_4$
is hyper-K\"ahler.

 Next let us turn to examine the geometry of $X_6$. Using
 (\ref{su2curv}), one can see that  the Riemann curvature of $X_6$ is
 \bea
R_{a_1a_2, b_1 b_2}=  \tfrac{1}{4} G_{[a_1| b_1 c} G^*_{a_2] b_2}{}^c - \tfrac{1}{4} G_{[a_1| b_2 c} G^*_{a_2] b_1}{}^c
    + 2 F_{[a_1| b_1 c m_1 m_2} F_{a_2] b_2}{}^{c m_1 m_2} \,.
 \eea
Moreover, (\ref{su2curv}) and the Bianchi identities imply that $F$ and $G$ are parallel
 \bea
  \nabla_A F_{b_1 b_2 b_3 m_1 m_2} = \nabla_A G_{b_1 b_2 b_3} = 0 \,.
 \eea
This in particular implies that the curvature of $X_6$ is parallel
and so $X_6$ is a symmetric space. Next observe that the fluxes can
be written as \bea &&F=\tfrac{1}{4} [H^1\wedge
\hat\omega_I+H^2\wedge \hat\omega_J+H^3\wedge \hat\omega_K] \,,\cr
&&G={\rm Re} \,G+i{\rm Im}\, G=H^4+i H^5~, \eea where $H^s$,
$s=1,\dots,5$, are parallel 3-forms on $X_6$ and $\hat\omega_I$,
$\hat\omega_J$ and $\hat\omega_K$ are the K\"ahler forms associated
with the hyper-complex structure on $Y_4$. Furthermore, the
conditions (\ref{M3-constraint}) and (\ref{selfdf}) imply that $H^s$
are anti-self-dual three-forms on $X_6$. The remaining conditions
conditions in terms of $H^s$ can now be written as
%\bea
 % {H}^i_{M_1 M_2 M_3} & = & ( \text{Re}(G_{M_1 M_2 M_3}), \text{Im}(G_{M_1 M_2 M_3}), 4 \text{Re}(F_{M_1 M_2 M_3 3 4}), \notag \\
% &&  4 \text{Im}(F_{M_1 M_2 M_3 3 4}), 4 \text{Im}(F_{M_1 M_2 M_3 3 \bar 3})) \,,
 %\eea
 \bea
  R_{a_1 a_2, a_3 a_4} - \tfrac{1}{2} \sum_s H^s_{[a_1| a_3 b|} H^s_{a_2] a_4}{}^b = 0 \,, \qquad % \notag \\
  %&&
  \nabla_{a_1} H^s_{a_2 a_3 a_4} = H^{[s}_{a_1 [b_1 b_2} H^{r]}_{ b_3 b_4]a_2} = 0 \,.
 \eea
These conditions are precisely those that one finds for the maximally supersymmetric solutions
 of  $(2,0)$ supergravity in six dimensions \cite{jose}. In particular $X_6$ is a six-dimensional Lorentzian Lie group with
anti-self-dual structure constants. These groups have been classified in \cite{jose} and  they are locally
isometric to $\bR^{5,1}$, $AdS_3\times S^3$ and $CW_6(\l {\bf 1})$,
 and
 \bea H^s= v^s H~,
 \eea
 where $H$ are the
structure constants of $X_6$ and $v$, $v^2=1$, is a constant vector.
The maximally supersymmetric IIB $SU(2)$-backgrounds have been
summarized in the introduction.

\newsection{Maximal $Sp(2) \ltimes \mathbb{R}^8$-backgrounds}

\subsection{Supersymmetry conditions}

A basis in the space of the $Sp(2) \ltimes \mathbb{R}^8$-invariant
Majorana-Weyl spinors is
\bea
 \eta_1 = 1 + e_{1234} \,,~~~ \eta_2 = i (1 - e_{1234}) \,,~~~ \eta_3 = i (e_{12} + e_{34}) \,.
\eea To find the conditions that the Killing spinor equations of
appendix A impose on the geometry of spacetime, it is convenient to
split up the ten-dimensional frame indices into $A = (-,+,i)$, where
$i = (\a, \bar \a)$ and $\a = (1, \ldots, 4)$.

The algebraic Killing spinor equations (\ref{facalg}) imply that
\bea P_+=P_i=0~, \eea i.e. only $P_-$ is non-vanishing. In addition,
the algebraic and the gravitino Killing spinor equations imply that
\bea G_{-+i}=G_{+ij}=G_{ijk}=0~, \eea i.e. the only non-vanishing
components are $G=e^-\wedge L$, where $L=\tfrac{1}{2} L_{ij}
e^i\wedge e^j$. These components are in addition constrained as \bea
L^{\bf 5}=0~, \eea where we have used the decomposition of the space
of two-forms,  $\Lambda^2(\bR^8)=\mathfrak{sp}(2)\oplus
3\Lambda^2_{\bf 5}\oplus 3\Lambda^2_{\bf 1}$, under $Sp(2)=Spin(5)$.
Therefore, one can write that \bea G=e^-\wedge
(L^{\mathfrak{sp}(2)}+ \ell^r \omega_r) \,,\eea where \bea
&&\omega_1=\omega_I=-i\delta_{\a\bar\b} e^\a\wedge e^{\bar\b}~,~~~
\cr &&\omega_2=\omega_J={\rm Re}(\epsilon_{\a\b} e^\a\wedge
e^\b)~,~~~\omega_3=\omega_K=-{\rm Im} (\epsilon_{\a\b} e^\a\wedge
e^\b)~, \eea are the Hermitian forms generated by the quaternionic
endomorphisms $I,J$ and $K$, and $\ell^r$ are spacetime functions.
We follow the notation of \cite{glp}.

Next let us turn to the conditions on the $F$ fluxes. The gravitino
Killing spinor equations (\ref{facpar}) together with the
self-duality of $F$ imply that
\bea
 F_{i_1 \cdots i_5} = F_{+ i_1 \dots i_4 } = F_{-+ i_1 i_2 i_3} = 0 \,.
\eea Therefore one can write \bea F=e^-\wedge M~,~~~M= \tfrac{1}{4!}
M_{i_1\dots i_4} e^{i_1}\wedge \dots\wedge e^{i_4}~. \eea In
addition, the Killing spinor equations imply that \bea M^{\bf 5}=0~,
\eea where we have used the decomposition of self-dual 4-forms,
$\Lambda^{4+}(\bR^8)=\Lambda^{4+}_{\bf 14}\oplus 3\Lambda_{\bf
5}^{4+}\oplus 6 \Lambda^{4+}_{\bf 1}$, under $Sp(2)$
representations\footnote{Using $\mathfrak{sp}(2)=\mathfrak{so}(5)$,
$\Lambda_{\bf 14}$ can be identified with the traceless symmetric
representation $\tilde S^2(\bR^5)$. }. Therefore, one can write \bea
F=e^-\wedge (M^{\bf 14}+  m^{rs} \omega_r\wedge \omega_s)~, \eea
where $(m^{rs})$ is a  symmetric matrix of spacetime functions.

Furthermore, the gravitino Killing spinor equation (\ref{facpar}) imposes the conditions
\bea
\Omega_{A,+i} = 0~,~~~\Omega_{A, ij}^{\bf 5}=0~,
\la{geomsp2}
\eea
on the geometry of spacetime, where the restriction to the five-dimensional $Sp(2)$ representation is made in the $i,j$ indices. Therefore, one
can write that
\bea
\Omega_{A,ij}= \Omega_{A,ij}^{\mathfrak{sp}(2)}+\Omega^r_A (\omega_r)_{ij}~.
\la{geomsp22}
\eea
Using the above expressions for the fluxes and the geometry, the parallel transport
equation becomes
\bea
 && \partial_A \epsilon - \tfrac{1}{2} i Q_A \epsilon+ \tfrac{1}{2} \Omega_{A, -+}\epsilon +\tfrac{1}{4}\Omega_A^r(\omega_r)_{ij} \Gamma^{ij}\epsilon = 0 \,,~~~~~~A \neq -\,\,
\cr
 && \partial_- \epsilon - \tfrac{1}{2} i Q_- \epsilon + \tfrac{1}{2} \Omega_{-, -+}\epsilon +\tfrac{1}{4}\Omega_-^r(\omega_r)_{ij} \Gamma^{ij}\epsilon
+ \tfrac{i}{8} m^{rs} (\omega_r)_{ij} (\omega_s)_{kl}
\Gamma^{ijkl}\epsilon \cr &&~~~~~~~+ \tfrac{1}{8} \ell^r
(\omega_r)_{ij}\Gamma^{ij} C^*\epsilon=0
%
%\cr
% && + i (F_{- 1 \bar 1 2 \bar 2} + F_{- 12 \bar 3 \bar 4}) + i (F_{- 1 \bar 1 3 \bar 3}
% + F_{- 2 \bar 2 4 \bar 4} - F_{- 1 2 \bar 3 \bar 4}) \Gamma^{1 \bar 1 3 \bar 3}
% + \notag \\
 %&& + \tfrac{1}{2} i F_{- 1234} \Gamma^{1234} + \tfrac{1}{2} i F_{- \bar 1 \bar 2 \bar 3 \bar 4}
 %\Gamma^{\bar 1 \bar 2 \bar 3 \bar 4} + \tfrac{1}{2} i F_{- 12} \cont{\rho} (\Gamma^{12 3 \bar 3}
 %- \Gamma^{1 \bar 1 34}) + \tfrac{1}{2}
 % i F_{- \bar 1 \bar 2} \cont{\rho} (\Gamma^{\bar 1 \bar 2 3 \bar 3}
 %- \Gamma^{1 \bar 1 \bar 3 \bar 4})  + \notag \\
  \la{parsp2}
\eea The components $L^{\mathfrak{sp}(2)}$,
$\Omega^{\mathfrak{sp}(2)}_{A}$ and $M^{{\bf 14}}$ do not appear in
the parallel transport equations and so the Killing spinor equations
do not constrain them further. The integrability condition of
(\ref{parsp2}) is the vanishing of the curvature of the associated
connection which depends on the fluxes. This leads to the flatness
conditions \bea
&&\partial_{[A}\Omega_{B],-+}=0~,~~~R_{AB}^r=0~,~~~\partial_{[A}
\hat Q_{B]}=0~, \cr &&\hat\nabla_A \ell^r=0~,~~~\hat\nabla_A
(m^{rs}- \tfrac{1}{3} \delta^{rs} {\rm tr}\, m)=0~, \la{intsp2} \eea
where $\hat\nabla$ is the connection and  $R_{AB}^r$ is the
curvature of the $\mathfrak{sp}(1)$ connection $\Omega^r$,
respectively,
 and
\bea \hat Q_A=Q_A~,~~~A\not=-~,~~~\hat Q_-=Q_-+\tfrac{20}{3} {\rm
tr}\, m~. \eea Notice that in this case $m^0= \tfrac{1}{3} {\rm
tr}\,m\,\sum_{r=1}^3 \omega_r\wedge \omega_r$, i.e.~it is
proportional to the $Sp(2)\cdot Sp(1)$-invariant four-form. It turns
out that the components $\Omega_{A,-+}, \Omega^r_A, \hat Q_A$ of the
connection can be set to zero with a gauge transformation in
$U(1)\times SO(1,1)\times Sp(1)\subset U(1)\times Spin(9,1)$. In
this gauge, one finds that the remaining conditions of
(\ref{parsp2}) together with $dP=0$ imply that \bea
\ell^r=\ell^r(v)~,~~~m^{rs}=m^{rs}(v)~,~~{\rm tr}\, m=-
\tfrac{3}{20} Q_-(v)~. \eea The expressions for the fluxes are
summarized in the introduction.

\subsection{Geometry and field equations}

In the lightcone frame $(e^-, e^+, e^i)$ which arises from the
description of spinors in terms of forms, the spacetime metric can
be written as $ds^2= 2 e^- e^++\delta_{ij} e^i e^j$. Choosing the
gauge $\Omega_{A,+-}=0$ and using  the conditions (\ref{geomsp2}),
one finds that  $\Omega_{A,+B}=0$. So the null vector field $X=e_+$
is parallel\footnote{There is a parallel null vector field
independent of the choice of gauge, i.e. if
$\Omega_{A,+-}=\partial_A f$, then $X= e^f e_+$ is parallel.}
\bea
\nabla X=0~.
\eea
The conditions (\ref{geomsp2}),  (\ref{geomsp22}) and (\ref{intsp2}) imply that the holonomy
of the Levi-Civita connection of the spacetime is
\bea
{\rm hol}(\nabla)\subseteq Sp(2)\ltimes\bR^8~.
\eea
Adapting coordinates along $X={\partial\over\partial u}$
and using that $X$ is rotation free, the spacetime metric can be written as
\bea
ds^2=2 dv (du+V dv+n_i e^i)+ \delta_{ij} e^i e^j~,~~~e^-=dv~,~~~e^+=du+V dv+n_i e^i~,
\la{metrsp2}
\eea
where all the components of the metric are independent of $u$ but
they may depend on $v$ and the remaining coordinates. Clearly the
spacetime is a pp-wave propagating on an eight-dimensional manifold
$Y_8$ given by $u,v={\rm const}$.  The metric of $Y_8$ is $d\tilde
s^2=\delta_{ij} e^i e^j$. It is straightforward to see that the
conditions on the geometry imply that the holonomy of the
Levi-Civita connection, $\tilde \nabla$, of $Y_8$ is contained in
$Sp(2)$, ${\rm hol}(\tilde \nabla)\subseteq Sp(2)$, i.e. $Y_8$ is a
hyper-K\"ahler manifold. Observe that the metric of $Y_8$ depends on
$v$ and so $v$ can be thought of as a deformation parameter of the
$Sp(2)$-structure.

Furthermore, one can use the torsion free conditions to compute the Levi-Civita connection of (\ref{metrsp2}). The result  has been presented in
(\ref{spinconnection}). In this case, the conditions on the geometry imply that $\Omega_{-,ij}$ take values in $\mathfrak{sp}(2)$.
The fluxes and conditions on the geometry are summarized in the introduction. The remaining cases with non-compact stability subgroup
can be analyzed in a similar way. Because of this, we shall not present all the details.

It is well known that the Killing spinor equations impose some of the supergravity field equations.
 So  it remains to find  the field equations that are not satisfied as consequence of the Killing spinor equations.
 Since the fluxes are null, the Bianchi identities reduce to $dP=dG=dF=0$. In addition after some investigation
 of the integrability equations of appendix A, one finds that if
 \bea
 E_{--}=0~,
 \eea
 then all the field equations are satisfied.
 This is the case for all maximally supersymmetric $G$-backgrounds for $G$ non-compact. Because of this,
 we shall not repeat this analysis in the other cases.

\newsection{Maximal $(SU(2) \times SU(2)) \ltimes \mathbb{R}^8$-backgrounds}

A basis in the space of the $(SU(2) \times SU(2)) \ltimes
\mathbb{R}^8$-invariant Majorana-Weyl spinors is
 \bea
  \eta_1 = 1 + e_{1234} \,,~~~ \eta_2 = i (1 - e_{1234}) \,,~~~ \eta_3 = e_{12} - e_{34} \,,~~~ \eta_4 = i (e_{12} + e_{34}) \,.
 \eea
To find the conditions that the Killing spinor equations of appendix
A impose on the geometry of spacetime, it is convenient to use
light-cone frame indices $A = (-,+,i)$ and split up  $i = (a, m)$
according to embedding $SO(4)\times SO(4)\subset SO(8)$. In
addition, we use holomorphic and anti-holomorphic indices,
$U(2)\times U(2)\subset SO(4)\times SO(4)$,  as
 $a = (\a, \bar \a)$, with $\a = (1,2)$, and $m = (\mu,
\bar \mu)$, with $\mu = (3,4)$.

The algebraic Killing spinor equations (\ref{facalg}) imply that
\bea
P_+=P_i=0~,
\eea
i.e. only $P_-$ is non-vanishing. In addition, the algebraic (\ref{facalg}) and gravitino (\ref{facpar})
Killing spinor equations imply that
 \bea
  G_{+A_1A_2}=G_{ijk}=0 \,.
 \eea
Therefore, the non-vanishing components of $G$ are \bea G=e^-\wedge
L~,~~~L=\tfrac{1}{2} L_{ij}\, e^i\wedge e^j~. \eea The Killing
spinor equations imply that
 \bea
 G_{-am}=0~.
 \eea
 Thus we find that
 \bea
 L=\tfrac{1}{2} (L_{ab} e^a\wedge e^b+L_{mn} e^m\wedge e^n)~.
 \eea
 Each of these components decomposes further under $SU(2)\subset SO(4)$ as
 $\Lambda^2(\bR^4)=3 \Lambda^2_{{\bf 1}}\oplus \mathfrak{su}(2)$. Therefore $L$ can be written
 as
 \bea
&&L=L^{\mathfrak{su}(2)\oplus \mathfrak{su}(2)}+ L^{{\rm inv}}~,
\cr
&&L^{{\rm inv}}=\ell^1\omega_1+ \ell^2 \omega_2+\ell^3 \chi_1
+\ell^4 \chi_2+\ell^5 \bar\chi_1+\ell^6\bar\chi_2\,,
\la{lsu2su2}
 \eea
 where $\omega_1=-i e^{1}\wedge  e^{\bar 1}-i e^{2}\wedge  e^{\bar 2}$ and $\chi= 2 e^1\wedge e^2$
 are the hermitian and holomorphic volume forms associated with $SU(2)\times \{1\}\subset SU(2)\times SU(2)$, respectively,
 and similarly for $\omega_2$ and $\chi_2$. Furthermore, $\ell^1,\dots, \ell^6$ are spacetime functions and the first
 component of $L$ takes values in $\mathfrak{su}(2)\oplus \mathfrak{su}(2)$ as indicated.

Next, let us turn to the conditions on the $F$ fluxes. Again, one
can show using the Killing spinor equations that the non-vanishing
components of $F$ can be written as \bea F= e^-\wedge M~,~~~M=
\tfrac{1}{4!} M_{ijkl} e^i\wedge e^j\wedge e^k\wedge e^l~. \eea The
gravitino Killing spinor equations (\ref{facpar}) together with the
self-duality of $F$ imply additional conditions on $M$. It turns out
that $M$ can be written as \bea M= m^0[\omega_1\wedge \omega_1+
\omega_2\wedge \omega_2]+ \tfrac{1}{4} M_{a_1 a_2 m_1m_2}
e^{a_1}\wedge e^{a_2}\wedge e^{m_1}\wedge e^{m_2}~. \eea The last
component is further restricted. Decomposing the last components of
$M$ in $SU(2)\times SU(2)$ representations, one can write that \bea
&&M= m^0[\omega_1\wedge \omega_1+ \omega_2\wedge \omega_2]+ \hat
M^{{\rm inv}}+ M^{({\bf 3}, {\bf 3})} \,,\cr && \hat M^{{\rm
inv}}=m^1\omega_1\wedge \omega_2+ {\rm Re}[m^2 \omega_1\wedge \chi_2
% \cr &&~~~~
+m^3 \omega_2\wedge \chi_1 +m^4 \chi_1\wedge \chi_2+m^5
\chi_1\wedge \bar\chi_2]~, \cr &&M^{({\bf 3}, {\bf 3})}=\tfrac{1}{4}
\tilde M_{\a\bar\b\m\bar\n}\, e^{\a}\wedge e^{\bar\b}\wedge
e^\m\wedge e^{\bar\n}~, \la{msu2su2} \eea where we have used the
decomposition $\Lambda^2(\bR^4)\otimes
\Lambda^2(\bR^4)=9\Lambda_{({\bf 1}, {\bf 1})}\oplus 3
\Lambda_{({\bf 1}, {\bf 3})}\oplus 3\Lambda_{({\bf 3}, {\bf
1})}\oplus \Lambda_{(\bf{3}, \bf{3})}$ under $SU(2)\times SU(2)$,
and $\tilde M$ traceless. Furthermore $m^0$ and $m^1$ are real and
$m^2, \dots, m^5$ are complex functions of spacetime, respectively.

The Killing spinor equation (\ref{facpar}) also restricts the geometry of spacetime. In particular, one finds that
 \bea
  && \Omega_{A, b m} = \Omega_{A,+ i} = 0 \,.
  \la{geoconsu2su2}
 \eea
 The  spin connection  can be written as
 as
 \bea
 \Omega_{A, ij}=\Omega^{\mathfrak{su}(2)\oplus \mathfrak{su}(2)}_{A, ij}+ \Omega_{A, ij}^{{\rm inv}}
 \eea
 in analogy with (\ref{lsu2su2}), where the decomposition is only in the $i,j$ indices. Using this,
 the parallel transport equation can be written as
 \bea
 &&  \partial_A \epsilon- \tfrac{1}{2} i Q_A \epsilon+ \tfrac{1}{2} \Omega_{A, -+}\epsilon +
  \tfrac{1}{4} \Omega^{{\rm inv}}_{A, ij} \Gamma^{ij} \epsilon = 0 \,, ~~~~~A\not=-
 \cr
&&  \partial_- \epsilon - \tfrac{1}{2} i Q_-\epsilon + \tfrac{1}{2} \Omega_{-, -+}\epsilon +
  \tfrac{1}{4} \Omega^{{\rm inv}}_{-, ij} \Gamma^{ij} \epsilon \cr
  &&
  %+ \tfrac{1}{2} i (F_{- 1 \bar 1 2 \bar 2} + F_{- 3 \bar 3 4 \bar 4})
  -2i m^0\epsilon+ \tfrac{1}{48} i \hat M^{{\rm inv}}_{ ijkl} \Gamma^{ijkl}\epsilon
  +  \tfrac{1}{8} L^{{\rm inv}}_{ ij} \Gamma^{ij} (C*) \epsilon= 0
  \,.
 \eea
 These parallel transport equations are independent of $\Omega^{\mathfrak{su}(2)\oplus \mathfrak{su}(2)}$,
 $L^{\mathfrak{su}(2)\oplus \mathfrak{su}(2)}$
 and $M^{({\bf 3}, {\bf 3})}$. So there are no further conditions on these components imposed by the
 Killing spinor equations. It remains to solve the above parallel transport equations. For this observe
 that the connection $\Omega^{{\rm inv}}$ takes values in $\mathfrak{su}(2)^\perp\oplus \mathfrak{su}(2)^\perp=
 \mathfrak{su}(2)\oplus \mathfrak{su}(2)$.
 This is because $\mathfrak{so}(4)=\Lambda^2(\bR^4)=\mathfrak{su}(2)\oplus \mathfrak{su}(2)$.
 The vanishing of the curvature implies that
 \bea
 &&\partial_{[A} \Omega_{B],-+}=0~,~~~~R^{{\rm inv}}=0~,~~~\partial_{[A} \hat Q_{B]}=0~,
 \cr
 &&\nabla^{inv}_A \hat M^{{\rm inv}}=\nabla^{{\rm inv}}_A L^{{\rm
 inv}}=0~,~~~A\not=-
 \la{parconsu2su2}
 \eea
 where
 \bea
 \hat Q_A= Q_A~,~~A\not=-~;~~~~\hat Q_-=Q_-+4 m^0~,
 \eea
and $\nabla^{{\rm inv}}$ is the covariant derivative and $R^{{\rm
inv}}$ is the curvature of  the connection $\Omega^{{\rm inv}}$,
respectively. As in the previous case, there is a local $U(1)\times
Spin(9,1)$ transformation to set $\Omega_{A,-+}=\hat Q_A=
\Omega^{{\rm inv}}_A=0$. In this gauge and using $dP=0$, we find
that (\ref{parconsu2su2}) imply that \bea m^0=-\tfrac{1}{4} Q_-(v)~,
\eea and  that the spacetime functions in (\ref{lsu2su2}) and
(\ref{msu2su2}) that determine $L^{{\rm inv}}$ and $M^{{\rm inv}}$
depend only on the $v$ coordinate. The description of the geometry
of spacetime is similar to that of the $Sp(2)\ltimes \bR^8$ case we
have already investigated. In particular, there is a null parallel
vector field $X$ and the holonomy of the Levi-Civita connection is
contained in $(SU(2)\times SU(2))\ltimes \bR^8$. Therefore the
spacetime is a pp-wave propagating in an eight-dimensional space
$Y_8$ which has holonomy\footnote{If $Y_8$ is compact and simply
connected, then it is a product $Y_8=M_1\times M_2$, where $M_1$ and
$M_2$ are four-dimensional hyper-K\"ahler manifolds.}
$Spin(4)=SU(2)\times SU(2)$. The results of our analysis have been
summarized in the introduction.

\newsection{Maximal $\bR^8$-backgrounds} \la{invspin}

To investigate the Killing spinor equations and the integrability conditions of
 the maximally  supersymmetric $\bR^8$-backgrounds, one needs the Majorana
 $\bR^8$-invariant spinors of IIB supergravity. A basis of the $\bR^8$-invariant spinors is
\bea \eta_1&=& 1+e_{1234}~,~~~\eta_2= i(1-e_{1234}) \cr \eta_3&=&
e_{12}-e_{34}~,~~~\eta_4= i (e_{12}+e_{34}) \cr \eta_5&=&
e_{13}+e_{24}~,~~~\eta_6= i (e_{13}-e_{24}) \cr \eta_7&=&
e_{23}-e_{14}~,~~~\eta_8= i (e_{23}+e_{14}) \eea Observe that these
spinors are characterized by the condition \bea \Gamma^-\eta=0~.
\eea
In this section we shall again use the the light-cone decomposition of the frame indices $A=
(-,+, i)$.
The algebraic Killing spinor equations (\ref{facalg}) and (\ref{facpar}) imply that
the non-vanishing components of $P$ and $G$ are
 \bea
  P = P_-\, e^- \,, ~~~~G=e^-\wedge L~,~~~~L=\tfrac{1}{2} L_{ij} e^i\wedge e^j~.
  \eea
  There are no further restrictions on $L$.
  Similarly, (\ref{facpar}) implies that the non-vanishing components
  of $F$ are
  \bea
  F=e^-\wedge M~,~~~M= \tfrac{1}{4!} M_{ijkl} e^i\wedge e^j\wedge e^k\wedge e^l~.
  \eea
  There are no further restrictions on $M$. The condition on the geometry in this case
  is
  \bea
  \Omega_{A, +i}=0
  \eea
  together with the parallel transport equations.
The parallel transport equation for $f$ now reads as follows. For $A \neq -$ we have
 \bea
 && \partial_A \epsilon - \tfrac{1}{2} i Q_A\epsilon + \tfrac{1}{2} \Omega_{A,-+}\epsilon
  + \tfrac{1}{4} \Omega_{A, ij} \Gamma^{ij}\epsilon = 0 \,, ~~~A\not=-
 \cr
  &&  \partial_- \epsilon - \tfrac{1}{2} i Q_-\epsilon + \tfrac{1}{2} \Omega_{-,-+}\epsilon
  + \tfrac{1}{4} \Omega_{-, ij} \Gamma^{ij}\epsilon
   + \tfrac{1}{8} L_{ij} \Gamma^{ij}
  C*\epsilon %+ \nonumber \\
  %&& \; \;
  + \tfrac{1}{48} i M_{ijkl} \Gamma^{ijkl}\epsilon = 0 \,.
 \eea
The connection $C$, see appendix A, takes values in $\mathfrak{gl}(16,\bR)=\mathfrak{gl}(8,\bR)\otimes \bH$. The
integrability conditions of the above parallel transport equations imply that
\bea
&&\partial_{[A} \Omega_{B],-+}=0~,~~~\partial_{[A}  Q_{B]}=0~,~~~R^{ij}_{AB}=0~,
\cr
&&\hat\nabla_A L_{ij}=0~,~~~\hat\nabla_A M_{ijkl}=0~,~~~A\not=-~,
\eea
where $\hat\nabla$ and $R^{ij}$ is the covariant derivative and the
curvature of $\Omega_{A,ij}$, respectively. A similar analysis to
the previous case reveals that in the gauge $Q_A=\Omega_{A,
-+}=\Omega_{A,ij}=0$, $L$ and $M$ depend  only on $v$. Our solutions
generalize those of \cite{pope} since they contain both $G$ and $F$
fluxes. Compare also our result with the eleven-dimensional
supergravity pp-wave solution of \cite{hull}. Generic backgrounds
preserve sixteen supersymmetries. However, for special choices of
fluxes the supersymmetry can be enhanced
  \cite{ bfhpwave, cvetic, hullb, bena}.
The results have been summarized in the introduction.

\newsection{pp-wave solutions with fluxes}

We have identified all maximally supersymmetric $G$-backgrounds, for
$G$ compact, up to a local isometry. It remains to extend this to
the cases where $G$ is non-compact. The torsion free condition
implies  that
\bea
  && \Omega_{i,j-} = e^I{}_{(i} \partial_v e_{j) I} +\tfrac{1}{2} (dn)_{ij} \,, \qquad \Omega_{-,-i} = \partial_i V - \partial_v n_I e^I{}_i \,, \nonumber \\
  && \Omega_{-,ij} = e^I{}_{[i} \partial_v e_{j] I} - \tfrac{1}{2} (dn)_{ij} \,. \label{spinconnection}
 \eea
 So to find the solutions in the non-compact case,  one has to find the most general
solution of (\ref{Einstein-eq}) and restrict  $ e^A\Omega_{A,ij}$ to
${\mathfrak {k}}$. This is a rather challenging problem in the case
that the fields depend on the coordinate $v$. However, the problem
is considerably simplified provided that the fields are taken to be
independent of $v$. In such a case,   the field equation reduces to
(\ref{Einstein-eq-trunc}) and $dn$ is required to take values in
${\mathfrak {k}}$. This equation is a Laplacian equation on the
eight-dimensional transverse space $Y_8$ for the
function\footnote{The function $\a$ can remain an arbitrary function
of $v$.} $V$ with a source term reminiscent of that of resolved
branes in \cite{gibbonspope}.
 The source term  depends on the fluxes and a rotation term
depending on $d\b$.
The simplest case is whenever the fluxes $F=G=0$ and $dn=0$. In this case, $V$ is a harmonic function of $Y_8$,
$\Box_8 V=0$. These are the standard type of pp-waves propagating on manifolds of holonomy $K$. Many such solutions
have been found by solving for $\a$. In particular, in the case $Y_8=\bR^8$, $V=\mu_0+\sum_i {\mu_i\over |y-y_i|^6}$.
A generalization of these solutions is to allow for the presence of fluxes. In particular, one can
take $L^{\mathfrak{k}}=\tilde M=0$ but $L^{\rm inv}=M^{\rm inv}\not=0$. In this case, the equation for $\a$
becomes
\bea
\Box_8 V=-2 \l^2~,
\eea
where $\l$ is a constant that depends on the coefficients of the invariant terms. This equation can be solved
in a variety of cases. For example if $Y_8=\bR^8$, then one can write
\bea
V=-A_{ij} y^i y^j+ B_i y^i+\mu_0+\sum_i {\mu_i\over |y-y_i|^6}~,~~~{\rm tr} A=\l^2~.
\eea
The additional term modifies the asymptotic behavior of the solution
 as $|y|\rightarrow \infty$ which is now a plane wave  instead of flat space.

 One can also construct examples with $dn\not=0$. In all these cases, $dn$ takes values in $\mathfrak{k}$.
 Solutions to these conditions are known in many cases. For example for $Y_8=\bR^8$, some solutions
 have been summarized in \cite{gip}.

 It is also possible to obtain under certain  conditions smooth solutions for $Y_8$ compact without boundary.
Integrating (\ref{Einstein-eq-trunc}) by parts and using
(\ref{fluxLM}), we find that
\bea
\int_{Y_8}\, d{\rm vol}\,\,[||dn||^2- 8 ||M||^2- ||L||^2  -4 ||P||^2 ] = 0~.
\eea
 This equation can be read as a condition for the cancelation
of field fluxes against angular momentum associated to the
spacetime. If $dn=0$ the above condition cannot be satisfied and
smooth solutions do not exist. The above condition can be written in
various ways. In particular using
 (\ref{fluxLMinv}) and the orthogonality in the decomposition of the fluxes, one finds that
\bea \int_{Y_8}\, d{\rm vol}\,\,[||dn||^2- 8 (||M^{\rm
inv}||^2+||\tilde M||^2)-
(||L^{\mathfrak{k}}||^2+||L^{\rm{inv}}||^2) -4 ||P||^2 ] = 0~.
\la{globcon} \eea In addition, in many cases (\ref{globcon}) depends
on the cohomology class $[dn]\in H^2(Y_8,\bR)$ and not on the
representative chosen. For example in the  Calabi-Yau case
(\ref{cysu4}),  the condition (\ref{globcon}) can be written as \bea
\int_{Y_8} [- \tfrac{1}{2} dn\wedge dn\wedge \omega^2-8( M^{\rm
inv}\wedge M^{\rm inv}+\tilde M\wedge \tilde M) +\tfrac{1}{2} \bar
L^{\mathfrak{k}}\wedge L^{\mathfrak{k}}\wedge \omega^2] \cr ~~~~~-[
4 \ell^* \ell+ 4 P_-^* P_- ] \,\, {\rm Vol}(Y_8)=0~. \eea To find a
solution, it remains to specify $dn$, $\tilde M$ and
$L^{\mathfrak{k}}$. The existence of these require additional
conditions, see e.g. \cite{green}. For example in the Calabi-Yau
case, the existence of $dn$ and  $L^{\mathfrak{k}}$ requires that
\bea \int_{Y_8}\, dn\wedge \omega^3=0~,~~~\int_{Y_8}\,
L^{\mathfrak{su}(4)} \wedge \omega^3=0~. \eea It is likely that
similar conditions are required for the remaining  cases. Many
examples can be constructed for $Y_8$ non-compact. However, this may
require case by case investigation.

\vskip 0.5cm

{\bf Acknowledgements} \vskip 0.1cm We would like to thank Jose
Figueroa-O'Farrill for helpful discussions. The research of D.R.~is
funded by the PPARC grant PPA/G/O/2002/00475 and U.G.~has a
postdoctoral fellowship funded by the Research Foundation
K.U.~Leuven.

\vskip 0.5cm

\setcounter{section}{0}

\appendix{Killing spinor and integrability conditions for maximal $G$-backgrounds}

The Killing spinors of maximally supersymmetric $G$-backgrounds
can be written as
\bea
\epsilon_i=\sum_jf_{ij}\eta_j~,~~~i,j=1,\dots,N_{\rm max}~,
\eea
where $\eta_p$,  $p\leq m$, are $G$-invariant Majorana spinors and $\eta_{m+p}=i\eta_p$,  $N_{\rm max}=2m$,
and $f=(f_{ij})$ is a $N_{\rm max}\times N_{\rm max}$ invertible
matrix with entries real spacetime functions. It has been shown in \cite{ggpg2, ggpr} that
the algebraic Killing spinor equations of IIB supergravity for the
 maximally supersymmetric $G$-backgrounds
can be written as \bea P_A \Gamma^A~\eta_p&=&0~,~~~p=1, \dots,m
\,,\cr \Gamma^{ABC} G_{ABC}~\eta_p&=&0~,~~~p=1, \dots,m~.
\la{facalg}
\eea Similarly, the gravitino Killing spinor equation can be
expressed as
\bea && \tfrac{1}{2}[ \sum_{j=1}^N(f^{-1}D_M f)_{pj}~
\eta_j-i\sum_{j=1}^N(f^{-1}D_M f)_{m+pj}~\eta_j]+\nabla_M \eta_p +
\tfrac{i}{48} \Gamma^{N_1\dots N_4 } \eta_p
 F_{N_1\dots N_4 M}=0\,,
 \cr
 && \sum_{j=1}^N(f^{-1}D_M f)_{pj}~ \eta_j+i\sum_{j=1}^N(f^{-1}D_M f)_{m+pj}~\eta_j+ \tfrac{1}{4} G_{MBC} \Gamma^{BC}\eta_p=0~,
 \la{facpar}
\eea
where we have set $N=N_{\rm max}$ for simplicity. In turn, these equations can be rewritten
as a set of algebraic conditions  on the fluxes and a parallel transport equation
associated with  the restriction of the supercovariant derivative along the bundle of Killing spinors.
The latter condition can be written as
$f^{-1} df +C=0$. This gives rise to the integrability condition $dC-C\wedge C=0$.

Sometimes it is helpful to express  (\ref{facpar}) in terms of the Killing spinors $\e$. This gives
\bea
  \partial_A\e - \tfrac{1}{2} i Q_A\e + \tfrac{1}{4} \Omega_{A,B_1 B_2}
  \Gamma^{B_1 B_2}\e + \tfrac{1}{48} i F_{A B_1 \cdots B_4}
  \Gamma^{B_1 \cdots B_4}\e + \tfrac{1}{8} G_{A B_1 B_2} \Gamma^{B_1
  B_2} C*\e=0 \,.
 \eea
 However in this form, the various terms that arise with different powers of gamma matrices
 are not linearly independent. The integrability condition is
 \bea
   && - \tfrac{1}{2} i ( \partial_{[A} Q_{B]} - \tfrac{1}{16} i G_{[A| D_1 D_2} G^*_{B]}{}^{D_1 D_2})\e
    \cr
  &&+ \tfrac{1}{2} ( \tfrac{1}{4} R_{AB C_1 C_2} - \tfrac{1}{12} F_{[A| C_1 D_1 \cdots D_3} F_{B] C_2}{}^{D_1 \cdots D_3}
     - \tfrac{1}{8} G_{[A| C_1 D} G^*_{B] C_2}{}^D ) \Gamma^{C_1 C_2}\e\cr
   &&+ \tfrac{1}{8} ( \nabla_{[A} G_{B] C_1
  C_2} - i Q_{[A} G_{B] C_1 C_2} - \tfrac{1}{2} i F_{[A| C_1 C_2 D_1
  D_2} G_{B]}{}^{D_1 D_2}) \Gamma^{C_1 C_2} C*\e\cr
   && +\tfrac{1}{48} i (\nabla_{[A} F_{B] C_1 \cdots C_4} - \tfrac{3}{4} i G_{[A| C_1 C_2} G^*_{B] C_3
  C_4})\Gamma^{C_1 \cdots C_4} \e \cr
   &&+ \tfrac{1}{144} F_{[A| C_1 \cdots C_3
  D} F_{B] C_4 \cdots C_6}{}^D \Gamma^{C_1 \cdots C_6}\e
  \cr
  && +\tfrac{1}{192} i F_{[A| C_1 \cdots
  C_4} G_{B] C_5 C_5}  \Gamma^{C_1 \cdots C_6} C*\e=0~.
 \eea

As we have already mentioned, the linear system that determines
the components of the field equations that are implied from the
Killing spinor equations simplifies for maximally supersymmetric G-backgrounds \cite{ggpr}.
In particular, one finds that
\bea
\big[\tfrac{1}{2}\Gamma^B
E_{AB}- i \G^{B_1B_2 B_3}\LF_{AB_1B_2B_3}\big]\eta_p&=&0\,,
\cr
\big[\G^B LG_{AB} -\G_A{}^{B_1\ldots B_4}BG_{B_1\ldots B_4}
\big]\eta_p&=&0\,,
\cr
\big[\tfrac{1}{2}\G^{AB}LG_{AB}+\G^{A_1\ldots
A_4}BG_{A_1\ldots
A_4}\big]\eta_p&=&0\,,
\cr
\big[LP+\G^{AB}BP_{AB}\big]\eta_p&=&0~,
~~~~p=1,\dots,m~,
\label{max-int-cond}
\eea
where the expressions for the field equations and our notation is explained in \cite{ggpr}.
We use this linear system to find the field equations that must be imposed in addition
to the Killing spinor equations for a supersymmetric configuration to be a solution
of the supergravity theory.


\begin{thebibliography}{00}
\addcontentsline{toc}{section}{References} \frenchspacing \small
\addtolength{\itemsep}{-4pt}

\bibitem{josew}
J.~M.~Figueroa-O'Farrill,
``Breaking the M-waves,''
Class.\ Quant.\ Grav.\  {\bf 17}, 2925 (2000)
[arXiv:hep-th/9904124].



\bibitem{ggpg2}
U.~Gran, J.~Gutowski and G.~Papadopoulos,
``The G(2) spinorial geometry of supersymmetric IIB backgrounds,''
Class.\ Quant.\ Grav.\  {\bf 23} (2006) 143
[arXiv:hep-th/0505074].

\bibitem{ggpr}
U.~Gran, J.~Gutowski, G.~Papadopoulos and D.~Roest,
``Systematics of IIB spinorial geometry,''
Class.\ Quant.\ Grav.\  {\bf 23} (2006) 1617
[arXiv:hep-th/0507087].

\bibitem{west}
J.~H.~Schwarz and P.~C.~West,
``Symmetries And Transformations Of Chiral N=2 D = 10 Supergravity,''
Phys.\ Lett.\ B {\bf 126} (1983) 301.


\bibitem{schwarz}
J.~H.~Schwarz, ``Covariant Field Equations Of Chiral N=2 D = 10
Supergravity,'' Nucl.\ Phys.\ B {\bf 226} (1983) 269.


\bibitem{howe}
P.~S.~Howe and P.~C.~West, ``The Complete N=2, D = 10
Supergravity,'' Nucl.\ Phys.\ B {\bf 238} (1984) 181.





\bibitem{ffp}
J.~Figueroa-O'Farrill and G.~Papadopoulos,
``Maximally supersymmetric solutions of ten- and eleven-dimensional
supergravities,''
JHEP {\bf 0303} (2003) 048
[arXiv:hep-th/0211089].


``Pluecker-type relations for orthogonal planes,''
arXiv:math.ag/0211170.


\bibitem{bfhpwave}
M.~Blau, J.~Figueroa-O'Farrill, C.~Hull and G.~Papadopoulos,
``A new maximally supersymmetric background of IIB superstring theory,''
JHEP {\bf 0201} (2002) 047
[arXiv:hep-th/0110242].

\bi{cahen}
M.~ Cahen and N.~Wallach, ``Lorentzian Symmetric Spaces'', Bull. Amm. Math. Soc. {\bf 76} (1970) 585.



\bibitem{ggpm}
J.~Gillard, U.~Gran and G.~Papadopoulos,
``The spinorial geometry of supersymmetric backgrounds,''
Class.\ Quant.\ Grav.\  {\bf 22} (2005) 1033
[arXiv:hep-th/0410155].

U.~Gran, J.~Gutowski and G.~Papadopoulos,
``The spinorial geometry of supersymmetric IIB backgrounds,''
Class.\ Quant.\ Grav.\  {\bf 22} (2005) 2453
[arXiv:hep-th/0501177].


\bibitem{kgd4}
  J.~Kowalski-Glikman,
  ``Positive Energy Theorem And Vacuum States For The Einstein-Maxwell
  System,''
  Phys.\ Lett.\ B {\bf 150} (1985) 125.

  \bibitem{ortin}
  P.~Meessen and T.~Ortin,
  ``The supersymmetric configurations of N = 2, d = 4 supergravity coupled to
  vector supermultiplets,''
  arXiv:hep-th/0603099.

\bibitem{gmr}
  J.~B.~Gutowski, D.~Martelli and H.~S.~Reall,
  ``All supersymmetric solutions of minimal supergravity in six dimensions,''
  Class.\ Quant.\ Grav.\  {\bf 20} (2003) 5049
  [arXiv:hep-th/0306235].

\bibitem{jose}
  A.~Chamseddine, J.~Figueroa-O'Farrill and W.~Sabra,
  ``Supergravity vacua and Lorentzian Lie groups,''
  arXiv:hep-th/0306278.

\bibitem{glp}
U.~Gran, P.~Lohrmann and G.~Papadopoulos,
``The spinorial geometry of supersymmetric heterotic string backgrounds,''
JHEP {\bf 0602} (2006) 063
[arXiv:hep-th/0510176].


\bi{bat}
A.~Batrachenko and W.~Y.~Wen,
  ``Generalized holonomy of supergravities with 8 real supercharges,''
  Nucl.\ Phys.\ B {\bf 690} (2004) 331
  [arXiv:hep-th/0402141].


\bibitem{ross}
  O.~Madden and S.~F.~Ross,
  ``Quotients of anti-de Sitter space,''
  Phys.\ Rev.\ D {\bf 70} (2004) 026002
  [arXiv:hep-th/0401205].


\bibitem{bhfp}
M.~Blau, J.~Figueroa-O'Farrill, C.~Hull and G.~Papadopoulos,
``Penrose limits and maximal supersymmetry,''
Class.\ Quant.\ Grav.\  {\bf 19} (2002) L87
[arXiv:hep-th/0201081].

M.~Blau, J.~Figueroa-O'Farrill and G.~Papadopoulos,
``Penrose limits, supergravity and brane dynamics,''
Class.\ Quant.\ Grav.\  {\bf 19} (2002) 4753
[arXiv:hep-th/0202111].



\bibitem{verlinde}
  H.~Ooguri, C.~Vafa and E.~P.~Verlinde,
  ``Hartle-Hawking wave-function for flux compactifications,''
  Lett.\ Math.\ Phys.\  {\bf 74} (2005) 311
  [arXiv:hep-th/0502211].

\bibitem{pope}
  M.~Cvetic, H.~Lu and C.~N.~Pope,
  ``Penrose limits, pp-waves and deformed M2-branes,''
  Phys.\ Rev.\ D {\bf 69} (2004) 046003
  [arXiv:hep-th/0203082].

\bibitem{hull}
  C.~M.~Hull,
  ``Exact pp Wave Solutions Of 11-Dimensional Supergravity,''
  Phys.\ Lett.\ B {\bf 139} (1984) 39.

   \bibitem{cvetic}
  M.~Cvetic, H.~Lu and C.~N.~Pope,
  ``M-theory pp-waves, Penrose limits and supernumerary supersymmetries,''
  Nucl.\ Phys.\ B {\bf 644} (2002) 65
  [arXiv:hep-th/0203229].

\bibitem{hullb}
  J.~P.~Gauntlett and C.~M.~Hull,
  ``pp-waves in 11 dimensions with extra supersymmetry,''
  JHEP {\bf 0206} (2002) 013
  [arXiv:hep-th/0203255].

\bibitem{gibbonspope}
  M.~Cvetic, G.~W.~Gibbons, H.~Lu and C.~N.~Pope,
  ``Ricci-flat metrics, harmonic forms and brane resolutions,''
  Commun.\ Math.\ Phys.\  {\bf 232} (2003) 457
  [arXiv:hep-th/0012011].

  \bibitem{bena}
  I.~Bena and R.~Roiban,
  ``Supergravity pp-wave solutions with 28 and 24 supercharges,''
  Phys.\ Rev.\ D {\bf 67} (2003) 125014
  [arXiv:hep-th/0206195].


\bibitem{green}
  M.~B.~Green, J.~H.~Schwarz and E.~Witten,
``Superstring Theory. Vol. 2: Loop Amplitudes, Anomalies And Phenomenology,'',
CUP (1987).


\bibitem{gip}
  G.~Papadopoulos,
  ``Rotating rotated branes,''
  JHEP {\bf 9904} (1999) 014
  [arXiv:hep-th/9902166].



%\bibitem{becker}
%K.~Becker and L.~S.~Tseng, ``A note on fluxes in six-dimensional
%string theory backgrounds,'' arXiv:hep-th/0410283.








\end{thebibliography}
\end{document}